\documentclass{emulateapj}
\slugcomment{Accepted by ApJ Letters}

\shorttitle{Super-Eddington X-ray bursts}
\shortauthors{Boutloukos, Miller, \& Lamb}

\begin{document}

\title{Super-Eddington Fluxes During Thermonuclear X-ray Bursts}

\author{Stratos Boutloukos\altaffilmark{1}, M. Coleman Miller\altaffilmark{1,2}, and Frederick K. Lamb\altaffilmark{3,4}}

\affil{{$^1$}Department of Astronomy and Maryland Astronomy Center for Theory and Computation, University of Maryland,\\College Park, MD 20742-2421, USA\\
{$^2$}Joint Space Science Institute (JSI), University of Maryland,
College Park, MD 20742-2421\\
{$^3$}Center for Theoretical Astrophysics and Department of Physics, University of Illinois at Urbana-Champaign,\\1110 West Green Street, Urbana, IL 61801-3080, USA}
\altaffiltext{4}{{$\!$}Also Department of Astronomy.}
\email{stratos@umd.edu}

\begin{abstract}
\noindent
It has been known for nearly three decades that the energy
spectra of thermonuclear X-ray bursts are often well-fit by
Planck functions with temperatures so high that they imply a
super-Eddington radiative flux at the emitting surface, even
during portions of bursts when there is no evidence of
photospheric radius expansion. This apparent inconsistency is
usually set aside by assuming that the flux is actually
sub-Eddington and that the fitted temperature is so high because
the spectrum has been distorted by the energy-dependent opacity
of the atmosphere. Here we show that the spectra predicted by
currently available conventional atmosphere models appear
incompatible with the highest-precision measurements of burst
spectra made using the \textit{Rossi X-ray Timing Explorer}, such
as during the \mbox{4U~1820$-$30} superburst and a long burst
from \mbox{GX~17$+$2}. In contrast, these measurements are
well-fit by Bose-Einstein spectra with high temperatures and
modest chemical potentials. Such spectra are very similar to
Planck spectra. They imply surface radiative fluxes more than a
factor of three larger than the Eddington flux. We find that
segments of many other bursts from many sources are well-fit by
similar Bose-Einstein spectra, suggesting that the radiative flux
at the emitting surface also exceeds the Eddington flux during
these segments. We suggest that burst spectra can closely
approximate Bose-Einstein spectra and have fluxes that exceed the
Eddington flux because they are formed by Comptonization in an
extended, low-density radiating gas supported by the outward
radiation force and confined by a tangled magnetic field.
\end{abstract}


\section{Introduction}
\label{sec:intro}

Type~I X-ray bursts (hereafter bursts) are produced by
thermonuclear burning of matter accumulated in the surface layers
of accreting neutron stars \citep{woos76,joss77,lamb78}. These
bursts have rise times ranging from a fraction of a second to a
few tens of seconds, durations ranging from about ten seconds to
several thousand seconds, recurrence times
$\sim\,$10$^3$--10$^6$~s, peak luminosities
$\sim\,$10$^{38}$~ergs~s$^{-1}$, and total energy releases
$\sim\,$10$^{39}$--10$^{42}$~ergs \citep{stro06}. The observed
X-ray flux typically increases by a factor of $\sim\,$10--100
during a burst. The properties of the large number of bursts that
have been observed using the \textit{Rossi X-ray Timing Explorer}
(\textit{RXTE}) have recently been summarized by \citet{gall08}.

Planck (blackbody) functions are often fit to the energy spectra
of bursts \citep{swan77, hoff77, gall08}. During some, the
temperature obtained from such fits drops and the derived
emitting area increases. These photospheric radius expansion
(PRE) bursts are thought to occur when the radiative flux through
the stellar atmosphere exceeds the Eddington critical flux,
creating an optically thick wind (see \citealt{gall08}). The
radiative flux is greater than the Eddington flux for any
realistic neutron star if the emission has a Planck spectrum with
a temperature measured at infinity $kT_\infty>2.0$~keV (see
\citealt{mars82} and Section~\ref{sec:max-temp}). Yet fits of
Planck functions to burst spectra frequently yield temperatures
substantially higher than this expected maximum, even during
times when there is no evidence of radius expansion.

Neutron stars are not blackbodies, and conventional model atmosphere
calculations show that they generally do not produce Planck
spectra (see, e.g., \citealt{lond84, lond86, made04, majc05}). In
conventional atmospheres, energy-dependent absorption and
scattering cause the spectrum to peak at an energy higher than
the peak of a Planck spectrum with the same effective
temperature. This effect led to widespread acceptance of the
hypothesis that the effective temperature is substantially
smaller than the temperature obtained by fitting a Planck
function to the burst spectrum and that the radiative flux is
sub-Eddington even when the fitted temperature exceeds 2.0~keV
(see, e.g., \citealt{ebis84} and \citealt{gall08}, section 2.2).

In contrast to conventional neutron star atmospheres, low-density
atmospheres extensive enough to fully Comptonize free-free and
cyclotron photons will produce Bose-Einstein spectra
$dN/dE\propto E^2/[\exp((E-\mu)/kT)-1]$ with chemical potentials
$\mu$ that satisfy $|\mu| \ll kT$ (see, e.g., \citealt{illa75}).
These spectra have almost the same shape and energy flux as a
Planck spectrum with the same temperature, because a Planck
spectrum is a Bose-Einstein spectrum with $\mu=0$. An important
aspect of Bose-Einstein spectra is that knowledge of the
radiation temperature and the chemical potential is sufficient to
determine the radiative flux from the emitting surface; knowledge
of the distance to the source or its luminosity is unnecessary.

Here we report analyses of \textit{RXTE} data taken during
high-temperature segments of a superburst from
\mbox{4U~1820$-$30} and a long burst from \mbox{GX~17$+$2}. Such
segments provide the best opportunity to test spectral models,
because the large number of counts collected allows the spectrum
to be measured with exceptionally high precision. We find that
the spectra predicted by currently available conventional
atmosphere models appear incompatible with the spectra during
these segments, whereas Bose-Einstein spectra fit these spectra
well. The fits give $|\mu| \la kT$ and values of $kT$
substantially greater than 2.0~keV, implying radiative fluxes at
the emitting surface more than a factor of three larger than the
Eddington flux. There is no evidence that the emitting surface is
expanded at these times. We find that the spectra of other bursts
from \mbox{4U~1820$-$30} and \mbox{GX~17$+$2} and bursts from
many other bursters are well-fit by similar Bose-Einstein
spectra, suggesting that the radiative flux also exceeds the
Eddington flux during these bursts.

\section{Maximum Blackbody Temperature}
\label{sec:max-temp}

The radiative flux from a neutron star atmosphere confined by
gravitation cannot exceed the Eddington flux. As explained in
Section~\ref{sec:intro}, the energy fluxes of Bose-Einstein
spectra with modest chemical potentials are very similar to the
fluxes of Planck spectra with the same temperature, so we can use
the Planck form as a proxy. The maximum allowed surface
temperature (measured at infinity) for emission with a Planck
spectrum from a star with mass $M$ and radius $R$ can be
determined by balancing the inward gravitational and outward
radiative accelerations at $R$. This maximum temperature is
\begin{equation}
\begin{array}{rl} \! kT_{\infty,{\rm max}} &=
4.60~{\rm keV}\left[(m/m_p)
(\sigma_T/\sigma)(M_\odot/M)\right]^{1/4}\\
&\quad\times\left(GM/Rc^2\right)^{1/2}(1+z)^{-3/4}\;, 
\end{array}
\label{eqn:Tmax}
\end{equation}
(see also \citealt{lewi93}, eq. (4.16)) where $k$ is the
Boltzmann constant, $m$ is the mass per nucleus, $m_p$ is the
proton mass, $\sigma_T$ is the Thomson scattering cross section,
$\sigma$ is the cross section per nucleus, $M_\odot$ is the solar
mass, and $1+z=(1-2GM/Rc^2)^{-1/2}$. Here $m$ and $\sigma$ are to
be evaluated at the photosphere and $z$ is the redshift from the
photosphere to infinity. Note that the maximum temperature is
independent of the distance to the source and the size of the
emitting area and depends only weakly on the mass of the star.

$kT_{\infty,{\rm max}}$ is largest for $GM/Rc^2=2/7$. This
largest value scales as $M^{-1/4}$. Assuming neutron star masses
are $\ge1.2~M_\odot$ (for comparison, the lowest mass determined
with high confidence is the $1.25~M_\odot$ mass of pulsar B in
PSR~J0737--3039; see \citealt{burg03}), $kT_{\rm max,H}=1.71$~keV
for an atmosphere of fully ionized hydrogen ($m=m_p$ and
$\sigma=\sigma_T$); $kT_{\rm max,He}=2.03$~keV for fully ionized
helium ($m=4m_p$ and $\sigma=2\sigma_T$). Similar results were
obtained by Marshall (1982; see also Hoshi 1981). $T_{\infty,{\rm
max}}$ depends on the composition of the atmosphere via
$m/\sigma\propto A/Z$, where $A$ and $Z$ are the atomic weight
and number; hence carbon or oxygen atmospheres have the same
$T_{\infty,{\rm max}}$ as a helium atmosphere. \textit{For the
rest of this paper we assume $kT_{\infty,{\rm max}} = 2.0$~keV.}

\section{Spectral Analysis and Results}
\label{sec:analysis-results}

All the data used in our analysis were obtained from the
\textit{RXTE} archive and were analyzed with FTOOLS version 6.8,
following the \textit{RXTE} cook book\footnote
{http://heasarc.gsfc.nasa.gov/docs/xte/recipes/cook\_book.html}
and using the recently updated \textit{RXTE} response generator
(v11.7) and calibration information. We usually subtracted the
average preburst emission during a 16~s interval preceding the
burst. We also constructed burst spectra without subtracting any
preburst emission, using \texttt{pcabackest} (version 3.8, also
recently improved) to estimate the purely instrumental
background. In all cases we considered only the energy range
3--27.5~keV (to concentrate on the thermal emission) and used the
data from all PCU layers.

\subsection{Long segments from \mbox{4U~1820$-$30} and
\mbox{GX~17$+$2}}

A superburst from \mbox{4U~1820$-$30} was observed on 1999
September 9 using \textit{RXTE}'s Proportional Counter Array
(PCA) with 16-s time resolution (Standard2 mode, 129 energy
channels). The data were studied by \citet{stro02}, who reported
best-fit Planck temperatures as high as 2.9~keV for about 800~s,
with no evidence of radius expansion during this interval. The
PCA spectrum at the burst peak shows an Fe~K$\alpha$ emission
line at zero redshift~\citep{stro02}, indicating that it is
produced outside the neutron star atmosphere.

We analyzed four 64-s segments of data from different parts of
this burst. The measured spectra of all of these segments are
similar and are well fit by Bose-Einstein spectra, with best-fit
temperatures ranging from 2.0~keV to 2.9~keV, but appear
inconsistent with the spectra predicted by currently available
conventional model atmospheres. Here we describe in detail our
analysis of the segment that began at MET=179460500.0,
$\sim\,$20~min after the start of the superburst.

\begin{figure}[t!]
\vspace{8pt}
\includegraphics[scale=0.45]{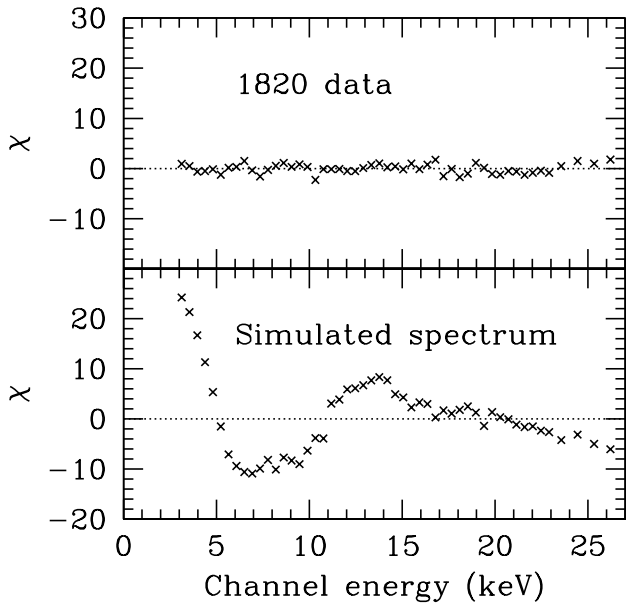}
\vspace{-20pt}
\caption{\label{fig-1820}
Residuals from fits of a Bose-Einstein spectrum to a spectrum of
the \mbox{4U~1820$-$30} system measured by \textit{RXTE} during a
superburst (top panel) and to the spectrum predicted by the
conventional H$+$He atmosphere model of \citet{made04} (bottom
panel). The fitting procedure is described in the text. In the
top panel, $\chi$ is the observed counts minus the Bose-Einstein
plus emission line counts, divided by the square root of the
observed counts; the best-fit temperature is 2.881~keV ($1\sigma$
range: 2.878~keV to 2.889~keV); the best-fit chemical potential
is $-0.48$~keV ($1\sigma$ range: $-0.53$~keV to $-0.40$~keV); and
$\chi^2/{\rm dof} = 43.8/44$. Setting $\mu=0$ (the Planck value)
gives a slightly inferior fit ($\chi^2/{\rm dof} = 53.5/45$). In
the bottom panel, $\chi$ is the simulated counts minus the
Bose-Einstein counts, divided by the square root of the simulated
counts. Comparison of the simulated spectrum with the
best-fitting Bose-Einstein spectral shape gives $\chi^2/{\rm
dof}=2918/48$. Results like these indicate that the emitted
spectrum is close to a Bose-Einstein spectrum, that the effective
temperature is $\approx3.0$~keV, and that the radiative flux at
the emitting surface exceeds the Eddington flux.}
\end{figure}

We first tested Bose-Einstein (adjustable $\mu$) and Planck
($\mu=0$) models for the spectra produced by burst atmospheres,
by fitting these models to the data. The Bose-Einstein spectrum
is not yet in XSPEC and we therefore used our own fitting
routines for it. Our routines reproduce the XSPEC results for
models that are in the library. We used the XSPEC routine
\texttt{bbody} to fit a Planck spectrum to the data. An external
emission line was included with both spectral models. Both
provide an excellent description of the measured spectrum (see,
e.g., Figure~\ref{fig-1820}). The implied radiative flux at the
emitting surface during this segment is at least four times
greater than the Eddington flux for any realistic neutron star.

We then investigated whether the spectra predicted by
conventional model atmospheres with sub-Eddington fluxes are
consistent with the spectrum we measured. No analytic
descriptions of these model spectra are available, so we used
standard methods (see, e.g., \citealt{cack09}) to construct the
PCA photon spectra they predict. We first redshifted the
published theoretical spectra by an amount appropriate to the
surface gravity of the model, using the APR equation of state
\citep{akma98}. We then constructed PCA count spectra using
\texttt{txt2xspec} (written by Randall Smith) and
\texttt{fakeit}. Finally, we normalized the count spectra to make
the total number of counts the same as in the spectrum measured
by the PCA.

Too few conventional atmosphere spectra have been published to be
able to fit them to the high-precision PCA spectra, so we
compared them with the Bose-Einstein spectral shape, which we
have shown provides an excellent description of the observed
spectrum, adjusting the shape to match the model spectra as
closely as possible. No external emission line was included,
because all these models describe the spectrum produced by the
burst atmosphere alone.

The conventional atmosphere models we studied are the $T_{\rm
eff} = 3.0\times 10^7$~K, $\log g=$14.8 H$+$He model of
\citet{made04} and the $T_{\rm eff}=2.0\times 10^7$~K, $\log
g=$14.1, 14.3, 14.5, and 14.7 solar composition models of
\citet{majc05}. These models span a substantial range of surface
effective temperatures $T_{\rm eff}$, surface gravities $g$, and
compositions. These models produce spectra that peak at an energy
higher than a Bose-Einstein spectrum with the same effective 
temperature but all have shapes that deviate systematically, 
similarly, and strongly ($\chi^2/{\rm dof} \ga 50$; see, e.g.,
Figure~\ref{fig-1820}) from the Bose-Einstein spectral shape that
describes the measured spectra. The deviations of the solar
composition models from the observed spectral shape are not
caused by the lines in these spectra: the energy response of the
PCA is much broader than these lines and the fit is not improved
significantly by excising them.

A long ($\sim\,$500-s) burst from \mbox{GX~17$+$2} was observed
on 1999 October 6 using the PCA. \cite{kuul02} have reported that
during a $\sim\,$100-s interval the spectrum of this burst is
well-fit by Planck spectra with $kT\ge 2.5$~keV and that there is
no evidence of radius expansion during this interval. We have
analyzed a 64-s segment of PCA data starting 10~s after the
beginning of this burst and confirm that Planck and Bose-Einstein
spectra fit this data well, whereas the available conventional
atmosphere spectra are again incompatible with the measured shape
of the spectrum ($\chi^2/{\rm dof}=1768/48$ for the conventional
H$+$He model atmosphere of \citealt{made04}).

\subsection{Shorter data segments}

In addition to analyzing four data segments from the
\mbox{4U~1820$-$30} superburst and a segment from the long
\mbox{GX~17$+$2} burst, we also analyzed shorter segments of PCA
data on many other, shorter bursts, to determine whether their
spectra are also well-fit by Bose-Einstein spectral models. These
data were taken using the Event mode, which provides 64 energy
channels. 

Using tables of the results obtained by \citet{gall08},
kindly provided by Duncan Galloway, we selected segments that are
well-fit by a Planck spectrum ($\chi^2/{\rm dof}<1.0$) and have a
best-fit temperature $T_{\infty}^{\rm best}$ well above 2.0~keV
($(kT_{\infty}^{\rm best}-2.0~{\rm keV})/(kT_{\infty}^{\rm best}
- kT_{\infty}^{\rm min})>5$, where $kT_{\infty}^{\rm min}$ is the
lower boundary of the 68.3\% temperature confidence interval;
$kT_{\infty}^{\rm best}$ and $kT_{\infty}^{\rm min}$ are given by
Galloway et al.). We found 1834 such segments, from 34 sources,
including \mbox{4U~1820$-$30} and \mbox{GX~17$+$2}; 4U~1728$-$34
is particularly prolific, with 556 such segments.

We again fitted Bose-Einstein and Planck spectra to the selected
data segments. For these segments, we included photoelectric
absorption using the XSPEC routine \texttt{phabs}. The results
listed in Table~\ref{tbl-fits} are typical of our results for all
these segments. The temperatures obtained by fitting
Bose-Einstein and Planck spectra to these intervals are
consistent with each other and are formally $>$10$\,\sigma$
higher than the 2.0~keV upper limit we established in Section~2,
implying that the fluxes during all these segments are
super-Eddington. These segments were selected as particularly
likely to have high temperatures, but they are representative of
the general burst population. \citet{kuul02} suggested that the
high fitted temperatures may be artifacts produced by subtracting
preburst emission. We therefore constructed spectra without
subtracting any preburst emission, but found that the
Bose-Einstein model also fit them and gave temperatures at least
as high as before.

\begin{table}[t]
\caption{Best-fitting parameters for 0.25-s segments of burst
data\tablenotemark{a}\label{tbl-fits}}
\begin{tabular}{cccc}
\tableline\tableline
\multicolumn{4}{c}{Aql~X-1, ObsID=60054-02-03-03, Starting MET=237409883.25}\\
\tableline
Model & kT (keV) & Chemical Potential & $\chi^2$/dof\\
\tableline
Bose-Einstein & $2.77\pm 0.05$ & $-2.5 \le \mu{\rm(keV)}\le -0.5$ & 15.6/26\\
Planck & $2.83\pm 0.03$ & {\rm ---} &15.6/27\\
\tableline\tableline
\multicolumn{4}{c}{4U~1702--429, ObsID=80033-01-19-04, Starting MET=333414491.50}\\
\tableline
Model & kT (keV) & Chemical Potential & $\chi^2$/dof\\
\tableline
Bose-Einstein & $3.07\pm 0.07$ & $-2.6 \le \mu{\rm(keV)}\le -0.5$ & 15.4/26\\
Planck & $3.04\pm 0.08$ & {\rm ---} & 16.6/27\\
\tableline\tableline
\multicolumn{4}{c}{EXO~1745--34, ObsID=50054-06-11-02, Starting MET=213117542.50}\\
\tableline
Model & kT (keV) & Chemical Potential & $\chi^2$/dof\\
\tableline
Bose-Einstein & $2.99\pm 0.06$ & $-2.5 \le \mu{\rm(keV)}\le -0.3$ & 21.4/26\\
Planck & $3.04\pm 0.05$ & {\rm ---} &22.3/27\\
\tableline
\end{tabular}
\tablenotetext{a}{All uncertainties are $1\sigma$. The $\mu$
range listed is the 68\% confidence interval. The fits are very
insensitive to the hydrogen column $N_H$, which we therefore do
not list.}
\end{table}

\subsection{Summary of results}

We have shown that Bose-Einstein spectral models with high
physical temperatures, modest chemical potentials, and
substantially super-Eddington fluxes at the emitting surface
provide excellent descriptions of high-precision measurements
of the spectrum near the peaks of the \mbox{4U~1820$-$30}
superburst and a long burst from
\mbox{GX~17$+$2}. The shapes of the spectra predicted by all the
currently available conventional model atmospheres appear
incompatible with the spectrum measured near the peaks of the
\mbox{4U~1820$-$30} superburst and the long burst from
\mbox{GX~17$+$2}. We have also found that the spectra of shorter
bursts from these two sources and many bursts from many other
bursters are well-fit by Bose-Einstein spectra with high
temperatures similar to the temperatures of the spectra that fit
the \mbox{4U~1820$-$30} superburst and the long \mbox{GX~17$+$2}
burst, suggesting that the radiative flux  also exceeds the
Eddington flux during these shorter bursts.

High spectral temperatures appear to be the rule rather than the
exception, particularly for PRE bursts. According to the data
tables of \citet{gall08}, which present fits of Planck spectra,
224 of 235 PRE bursts have at least one 0.25-second segment when
$kT_\infty>2.0$~keV and $\chi^2/{\rm dof}<1$; 488 of the 665
non-PRE bursts also have at least one such segment. If such high
spectral temperatures do indicate super-Eddington fluxes, these
results show that this phenomenon is widespread.

When the radiative flux falls below the Eddington flux, the burst
atmosphere may be supported by gas pressure gradients rather than
the radiation force. If so, the atmosphere will become more
dense, and the spectrum is likely to deviate from a Bose-Einstein
spectrum. \citet{gall08} have reported that burst spectra are
less likely to be described adequately by a Planck spectrum when
the flux is much less than the peak flux, results that hint at
this effect. It is possible that conventional model atmospheres
provide good descriptions of burst spectra when the flux is much
less that the Eddington flux.

\section{Discussion and Conclusions}
\label{sec:discussion-conclusions}

As discussed in Section~2, measurement of a Bose-Einstein
spectrum with a temperature greater than \mbox{$\sim\,$2~keV}
implies that the radiative flux at the emitting surface exceeds
the Eddington flux, independent of unknowns such as the distance
to the source, the radiating area on the star, the radius of the
star, and its surface redshift. The implied fluxes are accurate,
because the 2--60~keV bandpass of the PCA captures more than 95\%
of the flux of a 3.0~keV Bose-Einstein spectrum with $|\mu| \ll
kT$. We have found that intervals with temperatures greater than
$\sim\,$2~keV occur during most bursts, suggesting that the
radiative flux exceeds the Eddington flux during most bursts.
When combined with the flux profiles seen during PRE bursts from
some of these same stars, these results, and the small effective
areas inferred during high-temperature intervals, indicate that
most of the emission during these intervals comes from only a
fraction (in some cases $\sim\,$20\%) of the stellar surface.

These high temperatures and fluxes and small emitting areas raise
several important questions: How can the flux be super-Eddington
without producing a significant wind? What determines the maximum
flux from the emitting area, and how big is it? And how do these
results fit with evidence that the emitting surface is sometimes
far above the stellar surface?

We suggest that the radiative flux can exceed the Eddington flux
because the emitting gas is confined by a tangled stellar
magnetic field. The sudden nuclear energy release that produces a
burst creates a zone of turbulent convection at densities
$\sim\,$ 10$^{5-7}$~g~cm$^{-3}$ (see, e.g., \citealt{fush87}).
The convective energy flux is $F_t = \rho_t u_t^3$, where
$\rho_t$ is the density in the convection zone and $u_t$ is the
turbulent velocity there. The convection will amplify and tangle
the star's weak poloidal magnetic field until the tangled field
$B_t$ becomes strong enough to inhibit convection, which occurs
when $B_t^2/8\pi \approx \rho_t u_t^2$. The maximum value of
$B_t$ is $\approx (8\pi)^{1/2} \rho_t^{1/6} F_t^{1/3}$ and is
relatively insensitive to $\rho_t$ and $F_t$. For typical
densities in the convection zone and the highest energy fluxes
observed from the emitting surface, which are
$\sim\,$10$^{26}$~erg~cm$^{-2}$~s$^{-1}$, $B_t({\rm max})$ is
$\sim{\rm few}\times 10^{10}$~G, $\sim\,$10--100 times stronger
than the dipole components inferred from observations and
theoretical modeling (see \citealt{lamb08}).

The tangled field will be strong enough to confine the emitting
gas if its tension, $f_{\rm mag} \approx (1/4\pi)(B_t\cdot\nabla
B_t) \approx B_t^2/4\pi \ell_B$, exceeds the outward radiation
force, $f_{\rm rad} \approx (F_{\rm rad}/c) n_e\sigma$. Here
$\ell_B$ is the characteristic scale of the tangled field and
$n_e$ is the electron density in the radiative zone. Assuming
$\ell_B$ is no larger than the depth $\sim\,$10$^3$~cm of the
burning zone, a field $\sim\,$$B_t({\rm max})$ can confine the
atmosphere in the presence of a radiative flux $\approx
10^{26}$~erg~cm$^{-2}$~s$^{-1}$, which is the flux implied by an
effective temperature $\approx 3$~keV.

A neutron star atmosphere supported by a super-Eddington
radiative flux and confined by magnetic stresses is likely to be
more extended and have a lower density than a conventional
atmosphere supported by gas pressure and confined by gravity. In
a future paper (Lamb et al., in preparation), we show that such
an atmosphere naturally produces a Bose-Einstein photon spectrum
with $|\mu| \la kT$. Comptonization by the electrons in the
atmosphere drives the photon distribution close to a
Bose-Einstein distribution while weak free-free and cyclotron
emission drive the chemical potential to a small value (see,
e.g., \citealt{illa75}).

We expect that a region of very hot, confined gas will heat
adjacent areas of the stellar surface, which may not be confined
by a strong magnetic field. When these adjacent areas become hot
enough, they will expand vertically. If the product of the
radiative flux from the very hot, confined gas and its emitting
area exceeds the Eddington luminosity, adjacent gas will leave
the star as a wind, producing a PRE event. Hence the maximum
radiative luminosity will be approximately the Eddington
luminosity, just as in the conventional picture, even though heat
is flowing from below the atmosphere over only a fraction of the
stellar surface. The PRE will end when the \textit{luminosity} of
the very hot, confined gas falls below the Eddington luminosity,
even if the local \textit{flux} from this gas exceeds the
Eddington flux.

The high temperatures and radiative fluxes and small emitting
areas found here, which were first noted nearly three decades
ago, have important implications for efforts to determine neutron
star masses and radii using bursts. For example, it is often
assumed that during high-temperature intervals the entire stellar
surface emits exactly the Eddington flux. Our analysis of
spectral measurements made using \textit{RXTE} shows that these
assumptions must be reconsidered.

\acknowledgments
We thank Sudip Bhattacharyya, Duncan Galloway, Fotis Gavriil,
Ka-Ho Lo, and Tod Strohmayer for helpful advice. These results
are based on research supported by NSF grant AST0709015 and the
Fortner Chair at Illinois, and by NSF grant AST0708424 at
Maryland.

\clearpage
\end{document}